\documentclass[12pt]{article}
\usepackage{placeins}
\usepackage{amsmath,amssymb,amsfonts,amsthm,bm}    
\usepackage{lineno,hyperref}
\usepackage{algorithm2e}
\usepackage{graphicx}
\usepackage{pdfpages}
\usepackage[onehalfspacing]{setspace}
\usepackage[margin=1.5in]{geometry}
\usepackage{subcaption}
\usepackage{setspace}
\usepackage{placeins}
\usepackage{url}
\usepackage{xcolor}
\usepackage{mathtools}
\usepackage{titlesec}
\usepackage{natbib}

\titlelabel{\thetitle.\quad}

\begin{document}

	\def\spacingset#1{\renewcommand{\baselinestretch}%
		{#1}\small\normalsize} \spacingset{1}


		\title{\bf Regression-based Network Reconstruction with Nodal and Dyadic Covariates and Random Effects}
		\author{Michael Lebacher and G\"oran Kauermann\thanks{
				We would like to thank Samantha Cook and Kimmo Soram\"aki from Financial Network Analytics (\href{https://www.fna.fi/about}{www.fna.fi}) for providing and explaining the data and the problem to us.			
				The project was supported by the European Cooperation in Science and Technology [COST Action CA15109]. We also  acknowledge funding provided by the German Research Foundation (DFG) for the project  KA 1188/10-1.}\hspace{.2cm}\\
			Department of Statistics, Ludwig-Maximilians Universit\"at M\"unchen}

		\date{}	
		
		\maketitle

	\bigskip
	\begin{abstract}
	\noindent	Network (or matrix) reconstruction is a general problem which occurs if the margins of a matrix are given and the matrix entries need to be predicted. 
		In this paper we show that the predictions obtained from the iterative proportional fitting procedure (IPFP) or equivalently maximum entropy (ME) can be obtained by  restricted maximum likelihood estimation relying on augmented Lagrangian optimization. Based on the equivalence we extend the framework of network reconstruction towards regression by allowing for exogenous covariates and random heterogeneity effects. The proposed estimation approach is compared with different competing methods for network reconstruction and matrix estimation. Exemplary, we apply the approach to interbank lending data, provided by the Bank for International Settlement (BIS). This dataset provides full knowledge of the real network and is therefore suitable to evaluate the predictions of our approach. It is shown that the inclusion of exogenous information allows for superior predictions in terms of $L_1$ and $L_2$ errors. Additionally, the approach allows to obtain prediction intervals via bootstrap that can be used to quantify the uncertainty attached to the predictions. 
	\end{abstract}
	
	\noindent%
	\vfill
	
	\section{Introduction}
	The problem of how to obtain predictions for unknown entries of a matrix, given restrictions on the row and column sums is a problem that comes with many labels. Without a sharp distinction of names and fields, some non exhaustive examples for keywords that are related to very similar settings are \textsl {Network Tomography} and \textsl{Traffic Matrix Estimation} in Computer Sciences and Machine Learning (e.g.\ \citealp{vardi1996}, \citealp{coates2002}, \citealp{hazelton2010}, \citealp{airoldi2013}, \citealp{ZHOU2016220}), \textsl{Input-Output Analysis} in Economics (e.g.\ \citealp{bacharach1965}, \citealp{miller2009}), \textsl{Network Reconstruction} in Finance and Physics (e.g.\ \citealp{sheldon1998}, \citealp{squartini2011}, \citealp{mastrandrea2014}, \citealp{gandy2018}),  \textsl{Ecological Inference} in Political Sciences (e.g.\ \citealp{king2013}, \citealp{klima2016}),  \textsl{Matrix Balancing } in Operation Research (e.g.\ \citealp{schneider1990}) and many more.
	
	An old but nevertheless popular solution to problems of this kind is the so called \textsl{iterative proportional fitting procedure} (IPFP), firstly introduced by \cite{deming1940} as a mean to obtain consistency between sampled data and population-level information. In essence, this simple procedure iteratively adjusts the estimated entries until the row and column sums of the estimates match the desired ones. In the statistics literature, this procedure is frequently used as a tool to obtain maximum likelihood estimates for log-linear models in problems involving three-way and higher order tables (\citealp{fienberg1970}, \citealp{bishop1975},  \citealp{haberman1978,haberman1979}). Somewhat parallel, the empirical economics literature, concerned with the estimation of Input-Output matrices, proposed a very similar approach (\citealp{bacharach1970}), often called \textsl{RAS algorithm}. Here, the entries of the matrix must be consistent with the inputs and the outputs. The solution to the problem builds on the existence of a prior matrix that is iteratively transformed to a final matrix that is similar to the initial one but matches the input-output requirements. Although the intention is somewhat different, the algorithm is effectively identical to IPFP (\citealp{onuki2013}). The popularity of the procedure can also be explained by the fact that it provides a solution for the so called \textsl{maximum entropy} (ME) problem (\citealp{malvestuto1989}, \citealp{upper2011},  \citealp{elsinger2013}).  In Computer Sciences, flows within router networks are often estimated using \textsl{Raking} and so called \textsl{Gravity Models} (see \citealp{zhang2003}). Raking is in fact identical to IPFP and the latter can be interpreted as a special case of the former.
	
	In this paper, we propose an estimation algorithm that builds on  augmented Lagrangian optimization (\citealp{powell1969}, \citealp{hestenes1969}) and can provide the same predictions as IPFP but is flexible enough to be extended toward more general concepts. In particular we propose to include exogenous covariates and random effects to improve the predictions of the missing matrix entries. Furthermore, we compare our approach with competing models using real data. To do so, we look at an international financial network of claims and liabilities where we pretend that the inner part of the matrix is unknown. Since in the data at hand the full matrix is in fact available we can carry out  a competitive comparison with alternative routines. Note that commonly the inner part of the financial network remains unknown but finding good estimates for the matrix entries is essential for central banks and financial regulators. This is because it is a necessary prerequisite for evaluating systemic risk within the international banking system. See e.g.\ a very recent study by researchers from 15 different central banks (\citealp{anand2018}) where the question of how to estimate financial network linkages was identified as being crucial for contagion models and stress tests. Our proposal has therefore a direct practical contribution.
	
	The paper is structured as follows. In Section \ref{EM_model} we relate maximum entropy, maximum likelihood and IPFP. In Section \ref{constML} we introduce our model and discuss estimation and inference as well as potential extensions.
	After a short data description in Section \ref{application} we apply the approach, compare different models and give a small simulation study that shows properties of the estimator. Section \ref{conc} concludes our paper.

	\section{Modelling approach} \label{EM_model}
	\subsection{Notation}
	Our interest is in predicting non-negative, directed dyadic variables  $x_{ij}^t$ among $i,j=1,...,n$ observational units  at time points  $t=1,...,T$. The restriction to non-negative entries is henceforth referred to as \textsl{non-negativity constraint}.
	We do not allow for self-loops and leave elements $x_{ii}^t$ undefined. Hence, the number of unknown variables at each time point $t$ is given by $N=n(n-1)$. Let $\mathbf{x}^t=(x^t_{12},...,x^t_{1n},x^t_{21}...,x^t_{n(n-1)})^T$ be an $N$-dimensional column vector  and define 
	$\mathcal{I}=\{(i,j):i,j=1,...,n; i\neq j\}$
	as the corresponding ordered index set.
	We denote the $i$th row sum by $y^t_{i}=x^t_{i\bullet}=\sum_{j\neq i}x^t_{ij}$ and the $j$th column sum by $y^t_{n+ j}=x^t_{\bullet j}=\sum_{i\neq j}x^t_{ij}$.	Stacking the row and column sums, results in the $2n$-dimensional column vector $\mathbf{y}^t$. Furthermore, we define the known binary $(2n \times N)$ routing matrix $\mathbf{A}$ such that the linear relation
	\begin{equation}
	\label{margin}
	\mathbf{y}^t=\mathbf{A} \mathbf{x}^t\text{, for }t=1,...,T
	\end{equation}
	holds.
	Henceforth, we will refer to relation (\ref{margin}) as \textsl{marginal restrictions}. Furthermore, we denote each row of $\mathbf{A}$ by the row vector $\mathbf{A}_r=(a_{r1},...,a_{rN})$. Hence, we can represent the marginal restrictions row wise by 
	\begin{equation*}
	\mathbf{A}_r\mathbf{x}^t=y^t_r\text{, for }r=1,...,2n \text{ and }t=1,...,T.
	\end{equation*}
	Note that in cases where some elements of $\mathbf{y}^t$ are zero, the number of unknown variables to predict decreases and matrix $\mathbf{A}$ must be rearranged accordingly. In the following we will ignore this issue and suppress the time-superscript for ease of notation. Random variables and vectors are indicated by upper case letters, realizations as lower case letters.

	\subsection{Maximum entropy, iterative proportional fitting and maximum likelihood}  \label{proof}
	Besides the long known relation between maximum entropy (ME) and IPFP, there also exists an intimate relation between maximum entropy and maximum likelihood that is formalized for example by \citet{golan1996} and is known as the \textsl{Duality theorem}, see for example \citet{brown1986dual} and \citet{dudik2007}. Also in so-called configuration models (\citealp{squartini2011}, \citealp{mastrandrea2014}) the connection between maximum entropy and maximum likelihood is a central ingredient for network reconstruction.
	
	In the following (i) we rely on the work of \citet{golan1996}, \citet{squartini2011} and \citet{munoz2017} in order to briefly derive the ME-distribution in the given setting. (ii) After that, we show that IPFP indeed maximizes the  ME-distribution. (iii) Based on the first two results, we show that we can arrive at the same result as IPFP by constrained maximization of a likelihood where each matrix entry comes from an exponential distribution.\\
	\noindent\textsl{ (i) Maximum entropy distribution: }
	We formalize the problem by defining the Shannon entropy functional of the system as
	\begin{equation*}
	\label{entropy2}
	H[f]=-\int_{\mathcal{X}} f(\mathbf{x})\log (f(\mathbf{x}))\text{d}\mathbf{x},
	\end{equation*}
	where we make it explicit in the notation that the functional $H[f]$ takes the function $f$ as input. The support of $f$ is given by $\mathcal{X}\in \mathbb{R}_+^N$, ensuring the non-negativity constraint. Furthermore, we require that the density function $f: \mathcal{X} \to \mathbb{R}_+$ integrates to unity
	\begin{equation}
	\label{density}
	\int_{\mathcal{\mathcal{X}}}f(\mathbf{x})\text{d}\mathbf{x}=1.
	\end{equation}
	We denote the expectation of the random vector $\mathbf{X}$ by $\bm{\mu}$ and formulate the marginal restrictions in terms of linear restrictions on $\bm{\mu}$ which we specify as
	\begin{equation}
	\label{expected}
	\int_{\mathcal{X}}\mathbf{A}_r\mathbf{x}f(\mathbf{x})\text{d}\mathbf{x}=\mathbf{A}_r\bm{\mu}=y_r\text{ for }r=1,...,2n.
	\end{equation}
	Combining the constraints (\ref{density}) and (\ref{expected}) results  into the Lagrangian functional 
	\begin{equation}
	\label{F}
	\mathcal{L}[f]=-\int_{ \mathcal{X}} f(\mathbf{x})\log(f(\mathbf{x}))\text{d}\mathbf{x}-\lambda_0\bigg{(}\int_{ \mathcal{X}} f(\mathbf{x})\text{d}\mathbf{x}-1\bigg{)}-\sum_{r=1}^{2n}\lambda_r\bigg{(}\int_{ \mathcal{X}} \mathbf{A}_r\mathbf{x}f(\mathbf{x})\text{d}\mathbf{x} -y_r\bigg{)}
	\end{equation}
	with Lagrange multipliers $\lambda_r>0$ for $r=0,...,2n$.
	The solution can be found using the Euler-Lagrange equation (\citealp{dym2013}), stating that a functional of the form $\int_{ \mathcal{X}} L(\mathbf{x},f(\mathbf{x}),f'(\mathbf{x})) \text{d}\mathbf{x}$ is stationary (i.e.\ its first order derivative is zero) if
	\begin{equation}
	\label{eq:euler}
	\frac{\partial L}{\partial f}=\frac{\text{d} }{\text{d} \mathbf{x}}\frac{\partial L}{\partial f'}.
	\end{equation}
	If the Lagrangian functional does not depend on the derivative of $f(\cdot)$, we find the right hand side in equation (\ref{eq:euler}) to be zero so that  no derivative appears. For the Lagrangian functional  (\ref{F}) this provides 
	\begin{equation}
	\label{foc}
	-\log(\hat{f}(\mathbf{x}))-1-\lambda_0 - \sum_{r=1}^{2n}\lambda_r \mathbf{A}_r\mathbf{x}=0.
	\end{equation}
	Rearranging the terms in  (\ref{foc}) results in the maximum entropy distribution
	\begin{equation}
	\label{expfam}
	\hat{f}(\mathbf{x}) =\exp\bigg{\{}-\sum_{r=1}^{2n}\lambda_r \mathbf{A}_r\mathbf{x} -1-\lambda_0\bigg{\}}\text{, for } \mathbf{x}\in \mathcal{X}.
	\end{equation}
	In order to ensure  restriction (\ref{density})  we set $\exp(1+\lambda_0)=c(\bm{\lambda})$ where $\bm{\lambda}=(\lambda_1,...,\lambda_{2n})$ is the parameter vector and 
	\begin{equation*}
	c(\bm{\lambda})=\int_{ \mathcal{X}}\exp\bigg{\{}-\sum_{r=1}^{2n}\lambda_r \mathbf{A}_r\mathbf{x}\bigg{\}}\text{d}\mathbf{x},
	\end{equation*}
	where $\lambda_r>0$ for $r=1,...,2n$ ensures integration to a finite value.
	Taken together, this leads to the exponential family distribution
	\begin{equation}
	\label{expfam2}
	\hat{f}(\mathbf{x}) =\frac{1}{c(\bm{\lambda})}\exp\bigg{\{}-\sum_{r=1}^{2n}\lambda_r \mathbf{A}_r\mathbf{x}\bigg{\}}\text{, for } \mathbf{x}\in \mathcal{X}. 
	\end{equation}
	Apparently, the sufficient statistics in (\ref{expfam2}) result through
	\begin{equation*}
	\mathbf{A}_r\mathbf{x}= y_r\text{, for } r=1,...,2n
	\end{equation*}
	and hence, we can characterize the $N$ dimensional random variable $\mathbf{X}$ in terms of $2n$ parameters $\bm{\lambda}$. 
	Using (\ref{expfam2}), the second order condition results from
	\begin{equation*}
	-\frac{1}{\hat{f}(\mathbf{x})}<0\text{, } \forall \mathbf{x} \in \mathcal{X}
	\end{equation*}
	and ensures that $\hat{f}$ is indeed a maximizer.
	
	\noindent \textsl{(ii) IPFP and the maximum entropy distribution:} In order to solve for the parameters of the maximum entropy distribution we take the first derivative of the log-likelihood obtained from (\ref{expfam2}), i.e.\
	\begin{equation}
	\label{logl}
	\hat{\ell}(\bm{\lambda}) =-\log(c(\bm{\lambda})) -\sum_{r=1}^{2n}\lambda_ry_r .
	\end{equation}
	Since (\ref{expfam2}) is an exponential family distribution we can use the relation
	\begin{equation*}
	\frac{\partial \log( c(\bm{\lambda}))}{\partial \lambda_r}=-\mathbb{E}_{\bm{\lambda}}[\mathbf{A}_r\mathbf{x}]\text{, for }r=1,...,2n
	\end{equation*}
	and the maximum likelihood estimator  $\hat{\bm{\lambda}}$ results from the score equations
	\begin{equation}
	\label{score}
	1=\frac{\mathbf{A}_r\mathbf{x}}{\mathbb{E}_{\hat{\bm{\lambda}}}[\mathbf{A}_r\mathbf{x}]}=\frac{y_r}{\mathbb{E}_{\hat{\bm{\lambda}}}[\mathbf{A}_r\mathbf{x}]}\text{, for }r=1,...,2n.
	\end{equation}
	If we now multiply the left and the right hand side of (\ref{score}) by parameter $\lambda_r$ we get
	\begin{equation*}
	\lambda_r=\lambda_r\frac{y_r}{\mathbb{E}_{\hat{\bm{\lambda}}}[\mathbf{A}_r\mathbf{x}]}\text{, for }r=1,...,2n
	\end{equation*}
	and can solve the problem using fixed-point iteration (\citealp{dahmen2006}). That is we fix the right hand side to $\lambda_r^{k-1}$ and update the left side to $\lambda_r^{k}$ through
	\begin{equation}
	\label{ipf}
	\lambda_r^k=\lambda_r^{k-1}\frac{y_r}{\mathbb{E}_{\bm{\lambda}^{k-1}}[\mathbf{A}_r\mathbf{x}]}\text{, for }r=1,...,2n.
	\end{equation}
	But this is in fact iterative proportional fitting, a procedure that iteratively rescales the parameters until the estimates match the marginal constraints. Convergence is achieved when $\lambda_r^{k-1}=\lambda_r^{k}$, satisfying the score equations (\ref{score}).
	More generally, the log-likelihood (\ref{logl}) is monotonically  non-decreasing in each update step (\ref{ipf}) and convergence  of  (\ref{ipf})  is achieved only if the log-likelihood is maximized  (\citealp[Theorem 20.5]{koller2009}).
	
	\noindent \textsl{(iii) IPFP and constrained maximum likelihood:} If we re-sort the sufficient statistics and re-label the elements of $\bm{\lambda}$ we get
	\begin{equation*}
	\begin{split}
	\sum_{r=1}^{2n} \lambda_r \mathbf{A}_r\mathbf{x}=&\lambda_1(x_{12}+x_{13}+\dots+x_{1n})+\dots+\lambda_{2n}(x_{1n}+x_{2n}+\dots+x_{(n-1) n})\\
	=& \sum_{q=(q_1,q_2) \in \mathcal{I}} (\lambda_{q_1} + \lambda_{n+q_2})x_q
	= \sum_{q \in \mathcal{I}} \frac{x_q}{ \mu_q} 
	\end{split}
	\end{equation*}
	with $\mu_q=(\lambda_{q_1} + \lambda_{n+q_2})^{-1}$ for $q \in \mathcal{I}$. This leads to
	\begin{equation}
	c(\bm{\lambda})=\int_{\mathcal{X}} \exp\bigg{\{}-\sum_{q \in \mathcal{I}}\frac{x_q}{ \mu_q} \bigg{\}} \text{d} \mathbf{x}=\prod_{q\in\mathcal{I}}\mu_q,
	\end{equation}
	and  where with (\ref{score}) $\mathbf{A}\bm{\mu}=\mathbf{y}$.
	Hence, we can represent the whole system as the product of densities from exponentially distributed random variables $X_q$ for $q \in \mathcal{I}$. That is
	\begin{equation}
	\label{central}
	\hat{f} (\mathbf{x})=\exp\bigg{\{}-\sum_{q\in \mathcal{I}} \frac{x_q}{\mu_q} -\sum_{q\in\mathcal{I}}\log(\mu_q)\bigg{\}} =\prod_{q\in\mathcal{I}}\frac{1}{\mu_q}\exp\bigg{\{}-\frac{x_q}{\mu_q}\bigg{\}}
	\end{equation}
	with observed margins  $\mathbf{Ax}=\mathbf{A}\bm{\mu}=\mathbf{y}$ and $ x_q \geq 0$  $\forall$  $q\in\mathcal{I}$.

	\section{Maximum likelihood-based estimation strategy} \label{constML}
	\subsection{Parametrization and estimation}
	From result (\ref{central}) it follows that we can  use a distributional framework  in order to build a generalized regression model. We exemplify this with a model which includes a sender-effect denoted as $\bm{\delta}=(\delta_1,...,\delta_n)$ and a receiver-effect $\bm{\gamma}=(\gamma_1,...,\gamma_n)$. We stack the coefficients in a $2n$ parameter vector $\bm{\theta}=(\bm{\delta}^T,\bm{\gamma}^T)^T$ such that the following log-linear expectation results
	\begin{equation}
	\label{log_lin_mean}
	\mathbb{E}_{\bm{\theta}}[X_{ij}]=\mu_{ij}(\bm{\theta})=\exp(\delta_i + \gamma_j).
	\end{equation}
	Based on this structural assumption, we can now maximize the likelihood derived from  (\ref{central}) with respect to $\bm{\theta}$ subject to the observed values $\mathbf{Ax}=\bm{y}$ and the \textsl{moment condition} $\mathbf{A}\bm{\mu}(\bm{\theta})=\bm{y}$. In the given formulation, the  moment condition is linear in $\bm{\mu}(\bm{\theta})$ but not in $\bm{\theta}$. Consequently, the numerical solution to the problem might be burdensome.
	We therefore propose to use an iterative procedure that is somewhat similar to the Expectation Conditional Maximization (ECM, \citealp{meng1993}) algorithm, since it involves iteratively forming the expectation of $X_{ij}$ based on the previous parameter estimate (E-Step) and constrained maximization afterwards (M-Step). To be specific, define the $(N\times 2n)$ design matrix $\mathbf{Z}$, that contains indices for sender- and  receiver-effects. Matrix $\mathbf{Z}$ has rows $z_q$ indexed by $q\in\mathcal{I}$ where for $q=(i,j)$ we have the $i$-th and the$(n+j)$-th element of $z_q$ equal to $1$ and all other elements are equal to zero.
	Starting with an initial estimate $\bm{\theta}_0$ that satisfies $\mathbf{A}\bm{\mu}(\bm{\theta}_0)=\mathbf{y}$, we form the expectation of the log-likelihood 
	\begin{equation*}
	\label{emQ}
	Q(\bm{\theta}; \bm{\theta}_0)=\mathbb{E}_{\bm{\theta}_0}\bigg{[}\sum_{q\in \mathcal{I}}\bigg{(} -\mathbf{z}_q^T\bm{\theta} - \frac{X_{q}}{\mu_{q}(\bm{\theta})}  \bigg{)}\bigg{]}=\sum_{q\in \mathcal{I}}\bigg{(} -\mathbf{z}_q^T\bm{\theta} -\exp\{\mathbf{z}_q^T(\bm{\theta}_0-\bm{\theta}) \}  \bigg{)}.
	\end{equation*}
	Then, the maximization problem in the M-step is given by 
	\begin{equation}
	\label{problem}
	\max \limits_{\bm{\theta} \in \mathbb{R}^{2n}}  Q(\bm{\theta};\bm{\theta}_0) \text{ subject to } \mathbf{A}\bm{\mu}(\bm{\theta}) =\mathbf{y}.
	\end{equation}
	A suitable optimizer for non-linear constraints is available by the augmented Lagrangian (\citealp{hestenes1969}, \citealp{powell1969})
	\begin{equation}
	\label{auglag}
	\mathcal{L}(\bm{\theta};\bm{\xi}_k,\zeta,\bm{\theta}_k) = -Q(\bm{\theta};\bm{\theta}_k)-\bm{\xi}_k^T(\mathbf{A}\bm{\mu}(\bm{\theta}) -\mathbf{y}  ) + \frac{\zeta}{2} ||\mathbf{A}\bm{\mu}(\bm{\theta}) -\mathbf{y}    ||^2_2,
	\end{equation}
	with $\bm{\xi}_k$ and $\zeta$ being auxiliary parameters. The augmented Lagrangian method decomposes the constrained problem (\ref{problem}) into iteratively solving unconstrained problems. In each iteration we start with an initial parameter $\bm{\xi}_{k}$ in order to find the preliminary solution  	$\bm{\theta}_{k+1}$. Then, we update $\bm{\xi}_{k+1}=\bm{\xi}_{k}+\zeta(\mathbf{A}\bm{\mu}(\bm{\theta}_{k+1})-\mathbf{y})$ in order to increase the accuracy of the estimate. In case of slow convergence, also $\zeta$ can be increased.
	An implementation in  \texttt{R} is given by the package  \texttt{nloptr} by \citet{johnson2014}.
	
	
	
	
	\subsection{Confidence and prediction intervals}
	Considering the data entries as exponentially distributed allows for a quantification of the uncertainty of the estimates. We pursue this by bootstrapping (\citealp{efron1994}) here. Given a converged estimator $\hat{\bm{\theta}}$ we can draw for each matrix entry $X_{ij}$ from an exponential distribution with expectation $\mu_{ij}(\hat{\bm{\theta}})$ in order to obtain	 $B$ bootstrap samples $\mathbf{X}^*=(\mathbf{X}^*_{(1)},...,\mathbf{X}^*_{(B)})$. For each bootstrap sample  $\mathbf{X}^*_{(b)}$ we calculate the marginals $\mathbf{AX}^*_{(b)}=\mathbf{Y}^*_{(b)}$ and re-run the constrained estimation procedure resulting in  $B$ vectors of estimated means $\hat{\bm{\mu}}^*=(\hat{\bm{\mu}}^*_{(1)},...,\hat{\bm{\mu}}_{(B)}^*)$. Consequently, the moment condition $\mathbf{A}\hat{\bm{\mu}}^*_{(b)}=\mathbf{Y}^*_{(b)}$ holds for all mean estimates of the bootstrap and by model-construction, the expected marginal restrictions from the bootstrap sample match the observed ones:
	\begin{equation*}
	\mathbb{E}_{\hat{\bm{\theta}}}[\mathbf{A} \mathbf{X}_{(b)}^*]=\mathbf{A}\mathbb{E}_{\hat{\bm{\theta}}}[\mathbf{X}_{(b)}^*]=\mathbf{y}.
	\end{equation*}
	Based on the bootstrap estimates, we can easily derive confidence intervals for  $\mu_{ij}$ using the variability of $\hat{\mu}_{(b),ij}^*$ for $b=1,...,B$.
	Additionally, we define the prediction error as $e_{ij}=x_{ij}-\hat{\mu}_{ij}$ and construct prediction intervals for the unknown $x_{ij}$ based on the quantiles of the empirical distribution of
	\begin{equation*}
	\hat{\mu}_{ij}+e^*_{(b),ij}=\hat{\mu}_{ij}+x_{(b),ij}^*-\hat{\mu}^*_{(b),ij}\text{, for }b=1,...,B.
	\end{equation*}
	
	\subsection{Extensions with exogenous information and random effects}
	The regression framework allows to extend the model by including exogenous information. We consider again model (\ref{central}) and parametrize the expectation through
	\begin{equation}
	\label{log_lin_mean2}
	\mathbb{E}_{\bm{\theta}}[X_{ij}]=\mu_{ij}(\bm{\theta})=\exp(\delta_i + \gamma_j+ \tilde{\mathbf{z}}_{ij}^T\bm{\beta})=\exp(\mathbf{z}_{ij}^T\bm{\theta}),
	\end{equation}
	with $\delta_i$ and $\gamma_j$ again being the subject-specific sender- and receiver-effects. Furthermore, $\tilde{\mathbf{z}}_{ij}$ represents a $l$-dimensional covariate vector and $\bm{\beta}$ is the corresponding parameter vector.	We can use the augmented Lagrangian approach from above to estimate the $p=l+2n$ dimensional parameter vector $\bm{\theta}$. It is important to note here that only dyadic covariates have the potential to increase the predictive performance of the approach. If we only include subject-specific (monadic) information the expectation can be multiplicatively decomposed and the model collapses back to the IPFP model (\ref{log_lin_mean}). Thus is easily seen through
	\begin{equation*}
	\mu_{ij}(\bm{\theta})=\exp(\delta_i+\tilde{\mathbf{z}}_{i}^T\bm{\beta}_i + \gamma_j+ \tilde{\mathbf{z}}_{j}^T\bm{\beta}_j)=\exp(\tilde{\delta}_i+\tilde{\gamma}_j).
	\end{equation*}
	Nevertheless, the inclusion of subject-specific information may be valuable if it is the goal to forecast future networks based on new covariate information. This holds in particular in dynamic networks. We give an example for predictions based on lagged covariates in the next section.
	
	We can also easily add additional structure to model (\ref{log_lin_mean2}) and assume a distributional form for some or all coefficients. A simple extension arises if we assume random effects. This occurs by the inclusion of normally distributed sender- and receiver-effects:
	\begin{equation}
	\label{log_line2}
	\begin{split}
	(\bm{\delta},\bm{\gamma})^T&\sim \mathcal{N}_{2n}(\mathbf{0},\bm{\Sigma}(\bm{\vartheta})),
	\end{split}
	\end{equation} 
	where we take $\bm{\vartheta}$ as the vector of parameters that determines the covariance matrix of the random effects. The latter could be parametrized for example with $\bm{\vartheta}=(\sigma^2_\delta,\sigma^2_{\delta,\gamma}, \sigma^2_{\gamma})^T$ such that
	\begin{equation}
	\label{ranspec}
	\begin{pmatrix}
	\delta_i\\
	\gamma_j
	\end{pmatrix}\sim \mathcal{N}_2\left(\mathbf{0},\begin{pmatrix}
	\sigma^2_\delta & \sigma^2_{\delta,\gamma}   \\
	\sigma^2_{\delta,\gamma} & \sigma^2_{\gamma}  
	\end{pmatrix}\right)\text{, for }i,j=1,...,n\text{ and }i\neq j,
	\end{equation}
	where we assume separate variance components for the sender- and the receiver effects, respectively. 
	In order to fit the model, we follow a Laplace approximation estimation strategy similar to \citet{breslow1993}. Details are given in the Appendix \ref{raneff}.

	\section{Application}\label{application}
	
	\subsection{Data description}\label{data_des}
\begin{figure}[t!]
		\includegraphics[trim={0cm 0cm 0cm 0cm},clip,width=1\textwidth]{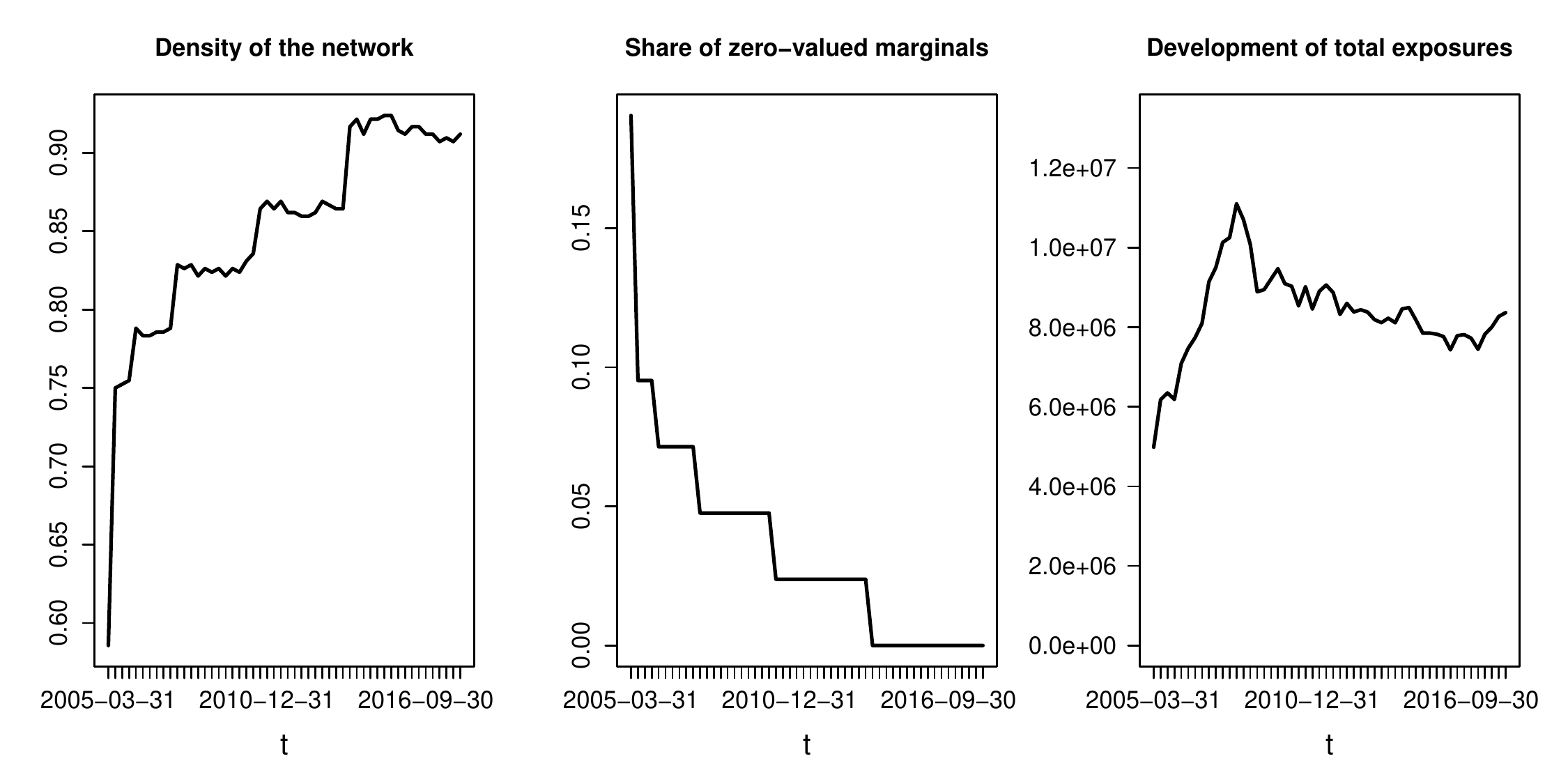}
		
		\caption{Density (left), share of zero-valued marginals (middle) and aggregated volume in million USD (right) of the network as quarterly time series.}
		\label{fig:density}
	\end{figure}

	\begin{table}[!t] \centering \small
		\begin{tabular}{@{\extracolsep{5pt}} clclcl} 
			\\[-1.8ex]\hline 
			\hline \\[-1.8ex] 
			AT & Austria & ES & Spain & JP & Japan \\ 
			AU & Australia & FI & Finland & KR & South Korea \\ 
			BE & Belgium & FR & France & NL & Netherlands \\ 
			CA & Canada & GB & United Kingdom & SE & Sweden \\ 
			CH & Switzerland & GR & Greece & TR & Turkey \\ 
			CL & Chile & IE & Ireland & TW & Taiwan \\ 
			DE & Germany & IT & Italy & US & United States of America \\ 
			\hline \\[-1.8ex] 
		\end{tabular} 
		\caption{Countries included in the analysis} 
		\label{countries} 
	\end{table} 
	The dataset under study is provided by the Bank for International Settlements (BIS) and freely available from their \href{https://www.bis.org/statistics/consstats.htm#}{homepage}. In general, the \textsl{locational banking statistics} (LBS) provide information about international banking activity by aggregating the financial activities (in million USD) to the country level. Within each country, the LBS accounts for outstanding claims (conceptualized as a valued dyad $x_{ij}$ that consists of all claims banks from country $i$ to banks of  country $j$) and liabilities of internationally active banks located in reporting countries (conceptualized as the reverse direction $x_{ji}$). We have selected the 21 most important countries (see Table \ref{countries}) for the time period from January 2005 to December 2017 as a quarterly series for the subsequent analysis. In Figure \ref{fig:density} the density of the network (number of existing edges relative to the number of possible edges) is shown on the left, the share of the zero-valued marginals in the middle and the development of the aggregated exposures on the right. Especially in the first years some marginals of the financial networks are zero and the corresponding matrix entries are therefore not included in the estimation problem. Correspondingly, it can be seen that most countries do have some claims and liabilities to other countries but especially in the beginning, many dyads $x_{ij}$ are zero valued. 
	
	Since it is plausible that financial interactions are related to the economic size of a country, we consider the annual \textsl{Gross Domestic Product} (GDP, in current USD Billions) as covariate. The data is provided for the years 2005-2017 by the International Monetary Fund on their \href{https://www.imf.org/en/Data}{homepage}. 
	Furthermore, there might be relationship between trade in commercial goods and financial transfers and we use data on \textsl{dyadic trade flows} (in current USD) between states as additional covariate. The data is available annually for the years 2005 to 2014 by the Correlates of War Project  \href{http://www.correlatesofwar.org/data-sets/bilateral-trade}{online}. We do not have  available information on trade for the years 2015, 2016 and 2017 and we therefore extrapolate the previous values using an autoregressive regression model. Apparently, by doing so we have covariate information which is subject to uncertainty. We ignore this issue subsequently. In order to have an example for  uninformative dyadic information, we use time-invariant data on the \textsl{dyadic distance} in kilometres between the capital cities of the countries under study (provided by  \citealp{gleditsch2013d}). Finally, in some matrix reconstruction problems, the matrix entries of previous time points become known after some time. Typically, lagged values are strongly correlated with the actual ones. We therefore also consider the matrix entries, lagged by one quarter as covariates. See Table \ref{covariates} for an overview of the variables, together with the overall correlation of the actual claims and the respective covariate. In the subsequent analysis we include all covariates in logarithmic form.
	
	\begin{table}[t!]\centering \small
		\resizebox{\textwidth}{!}{
			\begin{tabular}{clcr} 
				\hline 
				Variable&Description&Type & Correlation \\
				\hline \hline  
				$x_{ij}$& Claim from country $i$ to country $j$& dyad specific &$1.0000$\\
				\hline
				$gdp_i$& Gross Domestic Product of country $i$& node specific &$0.4716$\\
				$gdp_j$& Gross Domestic Product of country $j$& node specific&$0.1858$\\
				$trade_{ij}$& Bilateral trade flows of commercial goods& dyad specific &$0.4349$\\
				$dist_{ij}$&  Distance between the capital cities of countries $i$ and $j$& dyad specific &$-0.0953$\\
				$x_{ij}^{t-1}$& Lagged claim from country $i$ to country $j$& dyad specific &$0.9935$\\
				\hline \\[-1.8ex] 
			\end{tabular} 
		}
		\caption{Covariates used for the regression-based network-reconstruction} 
		\label{covariates} 
	\end{table}

	\subsection{Model performance}
	\begin{table}[t!]
		\resizebox{\textwidth}{!}{
			\begin{tabular}{clccccccccc} 
				\hline 
				&Method& Covariates&Rand. eff.&Model & overall $L_1$ & overall $L_2$ & average $L_1$ & SE & average $L_2$ & SE\\
				\hline \hline  
				1&IPFP & -&-& (\ref{ipf},\ref{log_lin_mean})	&$4\,204.212$ & $75.778$ & $80.850$ & $12.564$ &  $10.445$ & $1.168$ \\ 
				2&Regression &-&$\sigma_{\delta}^2, \sigma_{\gamma}^2,\sigma_{\delta,\gamma}^2$ & (\ref{log_lin_mean},\ref{ranspec})& $4\,204.212$ & $75.778$ & $80.850$ & $12.564$ & $10.445$ & $1.168$ \\  
				
				3&Regression& $gdp_i$, $gdp_j$, $trade_{ij}$&-& (\ref{log_lin_mean2})&{\bf3\,300.242}&   {\bf56.794}&   {\bf  63.466}&    $9.246$&    {\bf 7.802}&    $1.085$ \\ 
				4&Regression & $gdp_i$, $gdp_j$, $trade_{ij}$& $\sigma_{\delta}^2, \sigma_{\gamma}^2,\sigma_{\delta,\gamma}^2$ &(\ref{log_lin_mean2}, \ref{ranspec}&$3\,315.673$ &  $57.104$ &  $63.763$ &   {\bf 9.222}  &  $7.850$  &  {\bf1.052}\\
				5&Regression  &$gdp_i$, $gdp_j$, $dist_{ij}$&-& (\ref{log_lin_mean2})&$4\,884.728$ & $91.575$ & $93.937$ & $15.864$ & $12.565$ & $1.857$ \\ 
				6&Regression & $gdp_i$, $gdp_j$, $dist_{ij}$ &$\sigma_{\delta}^2, \sigma_{\gamma}^2,\sigma_{\delta,\gamma}^2$&(\ref{log_lin_mean2}, \ref{ranspec})& $4\,843.641$ &  $90.623$  & $93.147$&   $15.490$ &  $12.435$  &  $1.833$ \\ 
				\hline \\[-1.8ex] 
				7&	Regression  &$gdp_i$, $gdp_j$, $x_{ij}^{t-1}$&-& (\ref{log_lin_mean2})&{\bf2\,280.235} & {\bf41.805} & {\bf43.851} & $12.833$ & {\bf 5.591} & {\bf 1.549} \\ 
				8&	Regression & $gdp_i$, $gdp_j$, $x_{ij}^{t-1}$ &$\sigma_{\delta}^2, \sigma_{\gamma}^2,\sigma_{\delta,\gamma}^2$&(\ref{log_lin_mean2}, \ref{ranspec})& $2\,341.796$ & $43.483$ & $45.035$ & {\bf11.715} & $5.787$ & $1.710$ \\ 
				\hline \\[-1.8ex] 
			\end{tabular} 
		}
		\caption{Comparison of different regression models with the BIS Dataset. All values scaled by $100\,000$ and lowest values in bold.} 
		\label{comparison} 
	\end{table} 
	We evaluate the proposed models in terms of their $L_1$ and $L_2$ errors. The corresponding results are provided in Table \ref{comparison}. As a baseline specification, all models contain sender- and receiver effects. In the first row, we provide the maximum entropy model (\ref{log_lin_mean}) that coincides with the IPFP solution (\ref{ipf}). The second row shows model (\ref{log_lin_mean}) together with the random effects structure (\ref{ranspec}). In the third row, we provide the errors for model (\ref{log_lin_mean2}) where we included the covariates logarithmic GDP ($gdp_i$, $gdp_j$) as well as the logarithmic trade data ($trade_{ij}$). In row four, we use the same model as in row three but additionally added the random effects structure from (\ref{ranspec}). In rows five and six, the same models as in rows three and four are used but with logarithmic distance ($dist_{ij}$) instead of trade as dyadic explanatory variable. In the last two rows we consider models with lagged claims ($x_{ij}^{t-1}$) with and without random effects. This comparison might be somewhat unfair because of the strong correlation and because  it is not clear whether it can safely be assumed that such data is always available. Therefore, we have separated this specification from the other models.
	
	In the first four columns the different specifications together with the related equations are provided. Columns five and six show the aggregated errors over all 52 quarters and the last four columns show the errors averaged over all years together with their corresponding standard errors.
	\begin{figure}[t!]
		\includegraphics[trim={0cm 0.4cm 0cm 0cm},clip,width=1\textwidth]{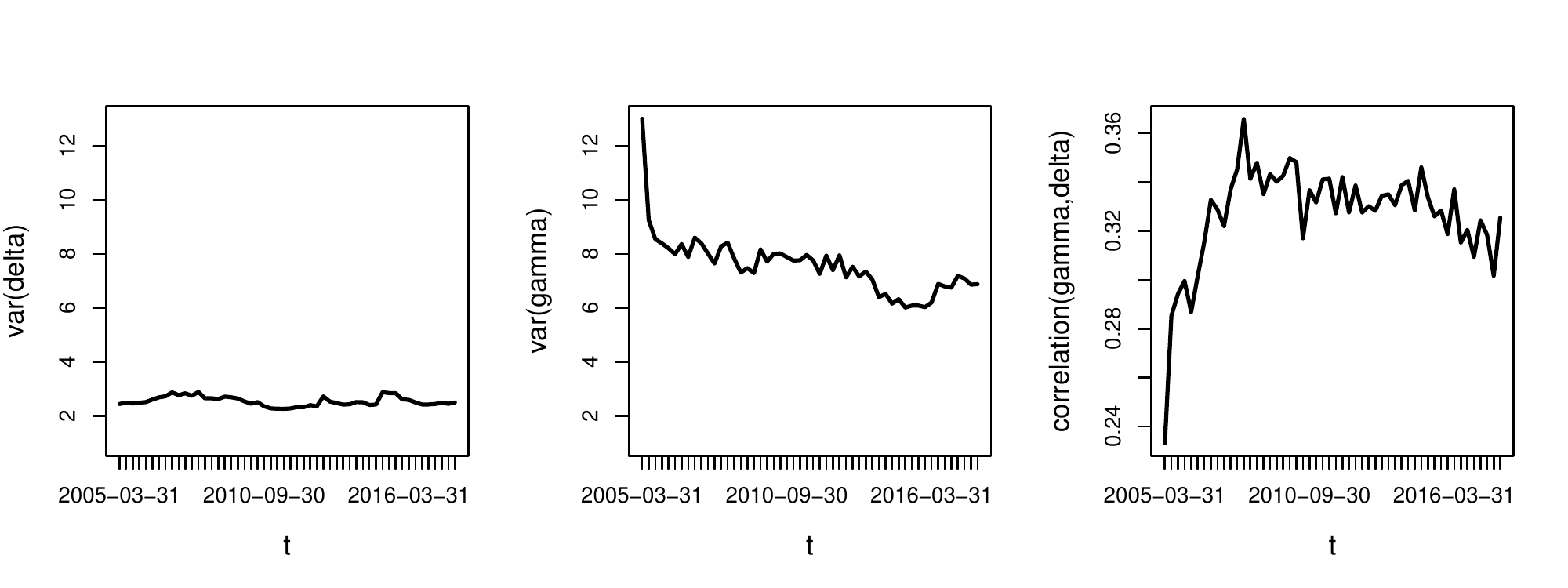}
		
		\caption{Quarterly time series of the estimated variances (models (\ref{log_lin_mean}) and (\ref{ranspec})) of the sender-effect ($\hat{\sigma}^2_{\delta}$) on the left, the receiver effect ($\hat{\sigma}^2_{\gamma}$) in the middle and the correlation between the sender- and the receiver effect ($\hat{\sigma}^2_{\delta,\gamma}/(\hat{\sigma}_{\gamma}\hat{\sigma}_{\delta})$) on the right.}
		\label{fig:sigmas}
	\end{figure}
	It can be seen that the first two models provide the same predictions and the inclusion of the random effects has no impact other than giving estimates for the variance of the sender- and receiver-effects as well as their correlation, shown in Figure \ref{fig:sigmas}. It becomes visible that the variation of the receiver-effect is much higher than the variation of the sender-effect which is almost constant. The correlation between the sender and the receiver effect is consistently positive and increases strongly within the first years.
	
\begin{figure}[t!]
		\centering		
		\begin{subfigure}{\textwidth}
			\centering			\includegraphics[trim={1cm 0.5cm 0cm 2cm},clip,width=\textwidth]{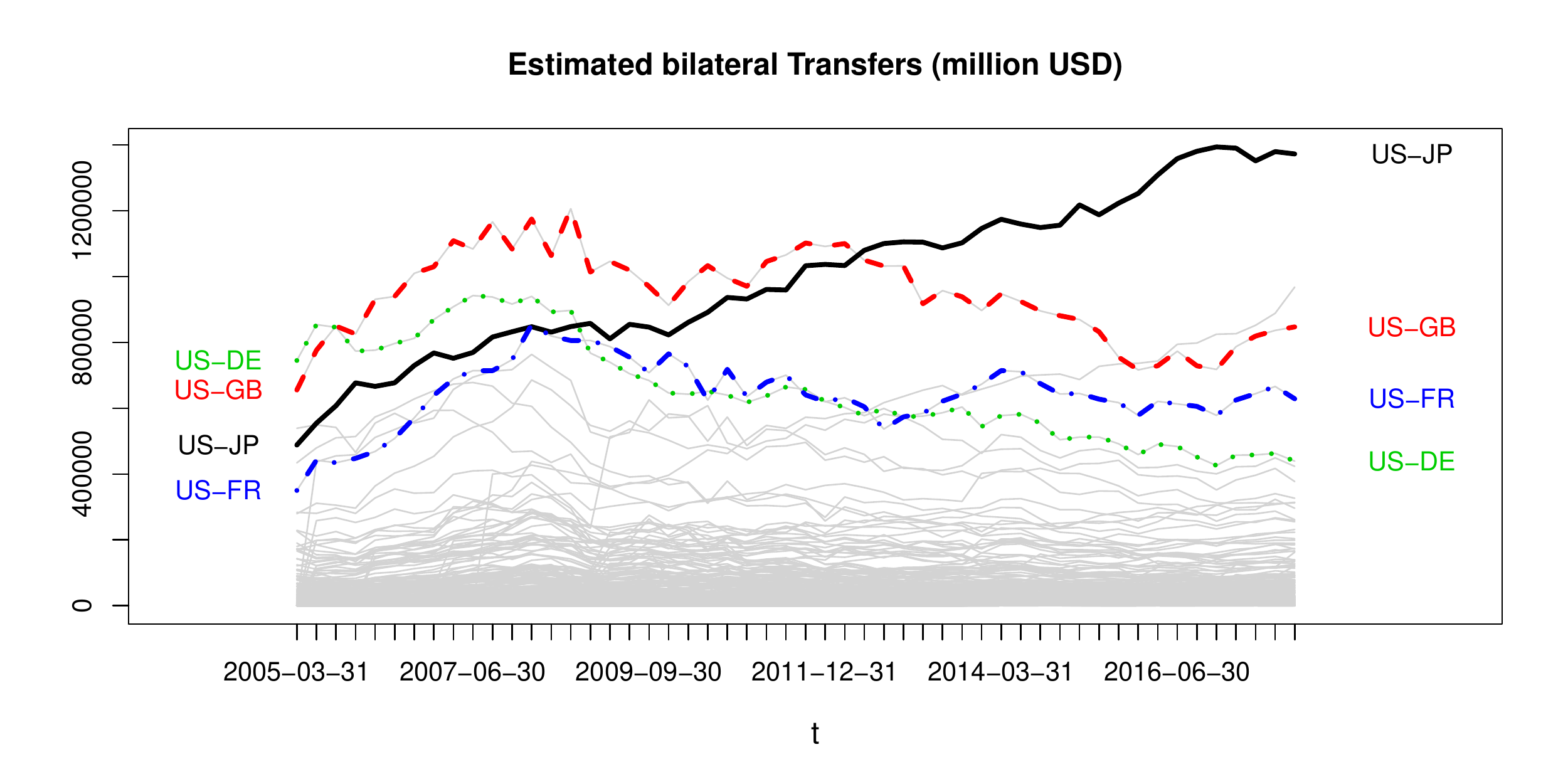}
		\end{subfigure}
		
		\caption{Fitted exposures, shown as quarterly series (52 time points) in million USD. All estimated with model (\ref{log_lin_mean2}) with $gdp_i$, $gdp_j$ and $trade_{ij}$ as covariates. Four dyads with the overall highest values are highlighted.}
		\label{fig:fv}
	\end{figure}

\begin{figure}[t!]
		\centering
		\begin{subfigure}{\textwidth}
			\centering			\includegraphics[trim={0cm 0.5cm 0cm 0cm},clip,width=\textwidth]{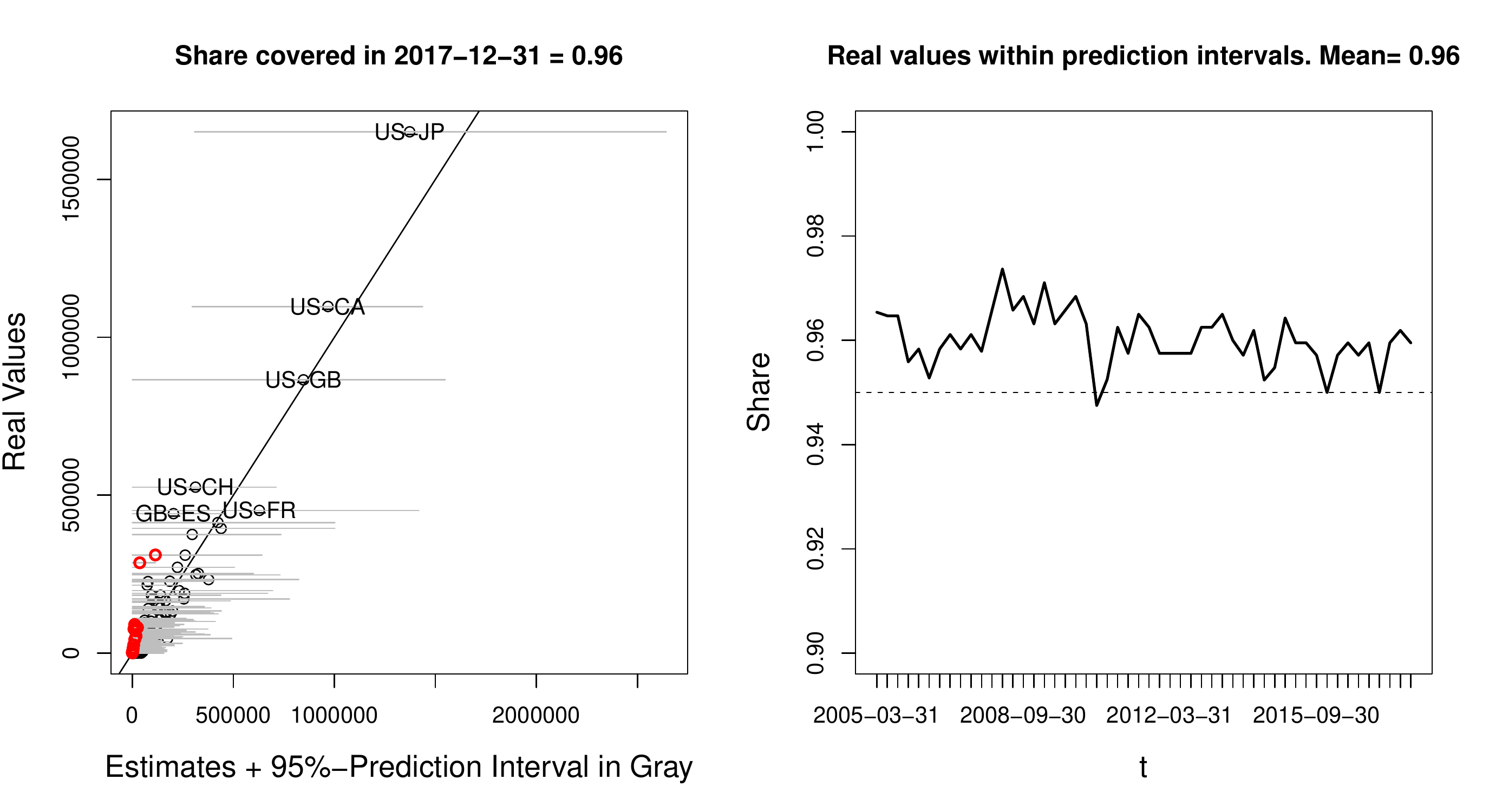}
		\end{subfigure}

		\caption{95\% Prediction intervals for the estimated means of model (\ref{log_lin_mean2}) with $gdp_i$, $gdp_j$ and $trade_{ij}$ as covariates, for the most recent network on the left. Prediction intervals in grey. Real values on the horizontal axis and estimated means on the vertical axis. Bisecting line in solid black. Predictions (not) covered by prediction intervals in black (red) circles. Share of real values within prediction intervals as a time series on the right.}
		\label{fig:summary2}
	\end{figure}

	Furthermore, Table \ref{comparison} shows that in the four models that include exogenous information (rows three to six) the extension towards the random effects structure has an impact on the predictive quality. It decreases in the model that includes the variable $trade_{ij}$ and increases in the one that includes $dist_{ij}$. Nevertheless, the models with and without random effects are rather close to each other and in fact they are statistically indistinguishable with respect to their $L_2$ differences. While the model with the covariate $dist_{ij}$ performs even worse than the IPFP solution, the model that includes $trade_{ij}$ but includes no random effects (row three) gives superior predictions relative to all other models in the upper part of Table \ref{comparison}. However, the two models that use the information on the lagged values give by far the best predictions. We nevertheless continue with the best model from the upper part of  Table \ref{comparison} lagged data are not necessarily available.
	\begin{figure}[t!]
		\centering
		\begin{subfigure}{\textwidth}
			\centering	\includegraphics[trim={0cm 0.5cm 0cm 2cm},clip,width=\textwidth]{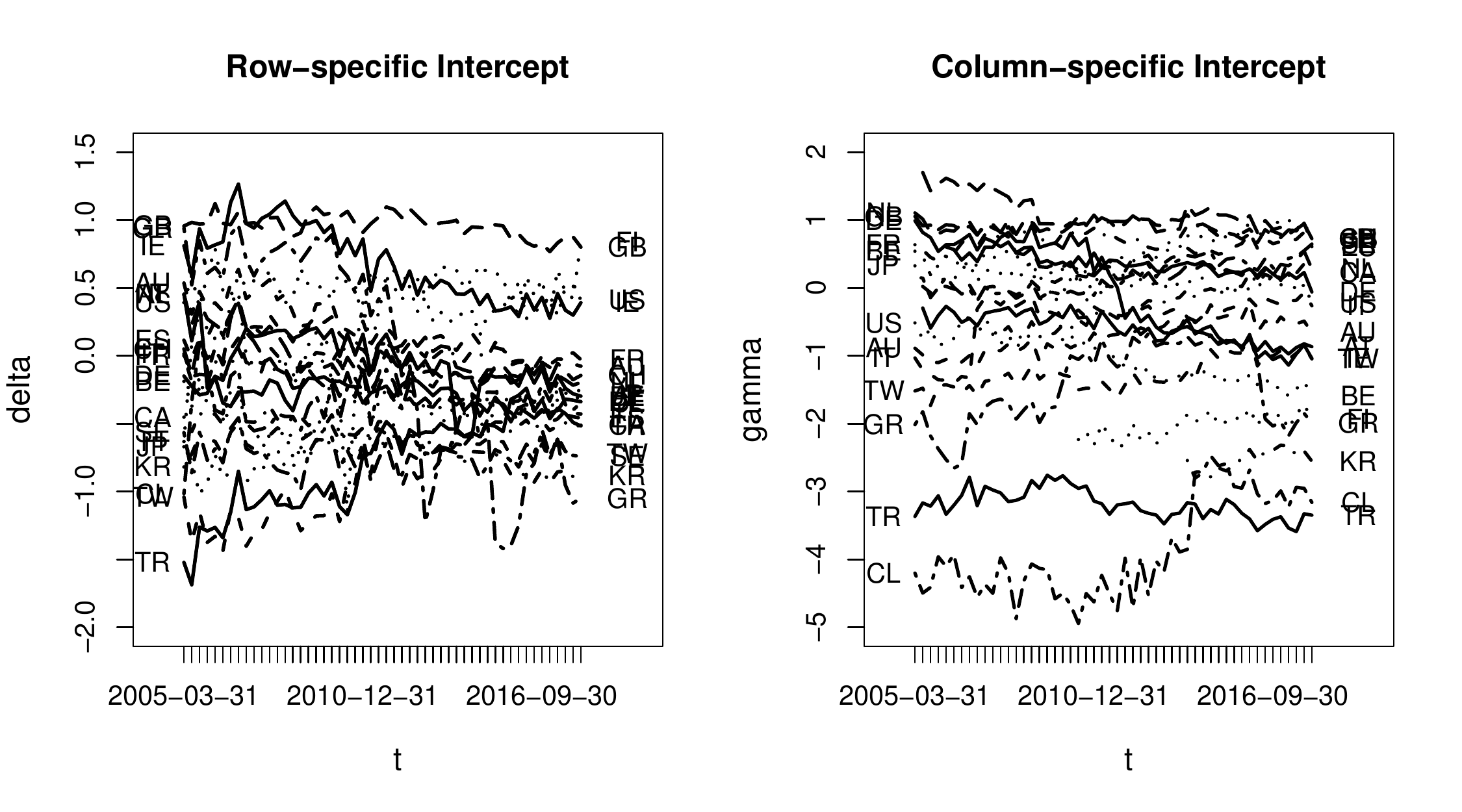}
		\end{subfigure}
		\begin{subfigure}{\textwidth}
			\centering		\includegraphics[trim={0cm 0.2cm 0cm 0cm},clip,width=\textwidth]{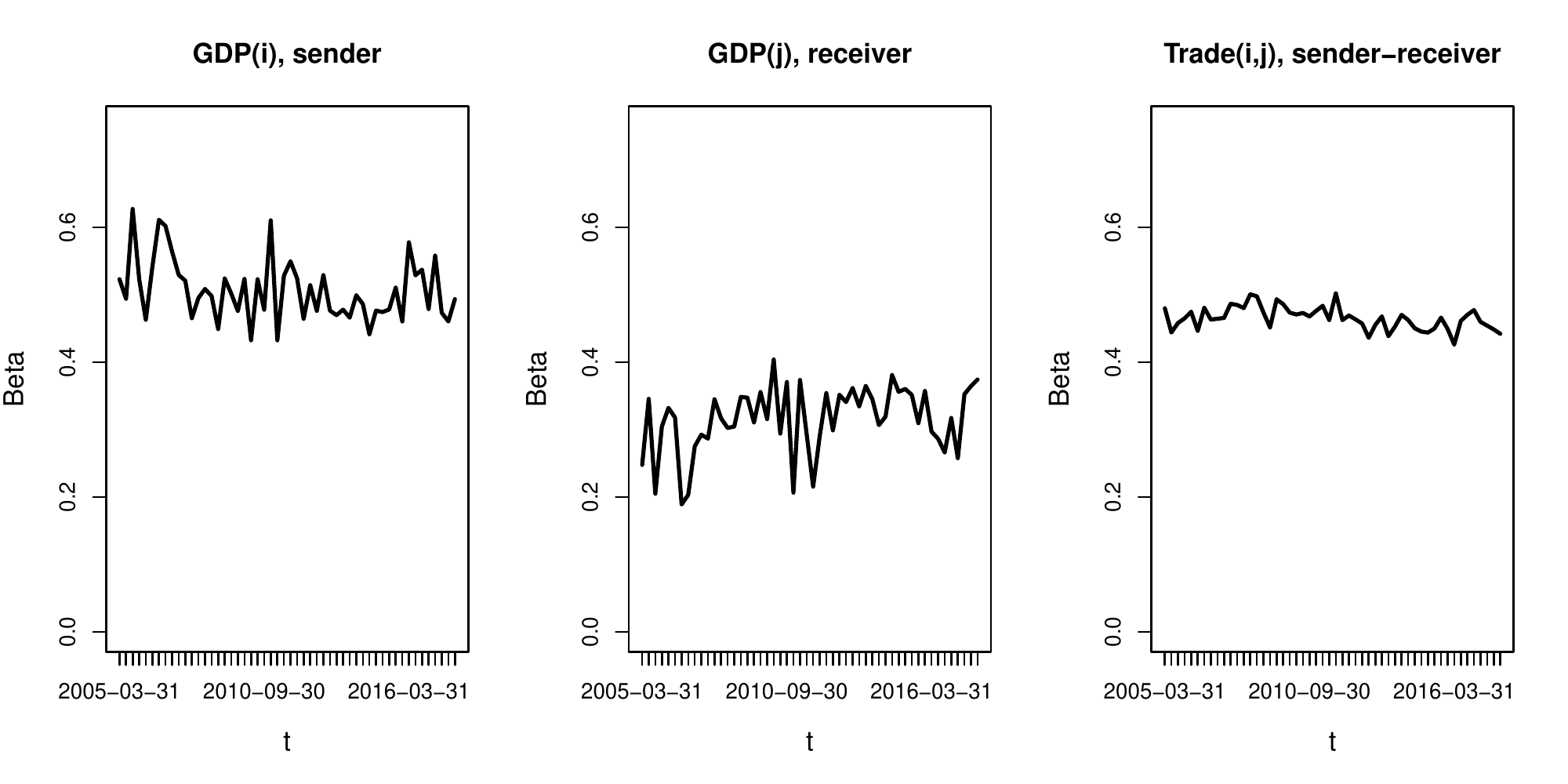}
		\end{subfigure}
		\caption{Estimated coefficients of model (\ref{log_lin_mean2}) with $gdp_i$, $gdp_j$ and $trade_{ij}$ as covariates. Time series of sender- ($\hat{\bm{\delta}}$, left) and receiver-effects ($\hat{\bm{\gamma}}$, right) in  the first row. Time series of estimated coefficients on exogenous covariates ($\hat{\bm{\beta}}$) in the second row.}
		\label{fig:summary1}
	\end{figure}
	The corresponding fitted values are provided as time series in Figure \ref{fig:fv} and in Figure \ref{fig:summary1} we provide the estimates for the coefficients of the model. In the first row, the estimated sender- ($\hat{\bm{\delta}}$, left) and receiver-effects ($\hat{\bm{\gamma}}$, right) are shown as a time series.  In the second row of Figure \ref{fig:summary1} the estimates for the coefficients on the exogenous covariates ($\hat{\bm{\beta}}$) can be seen. 
	The estimated coefficients provide the intuitive result that the claims from country $i$ to country $j$	increase with $gdp_i$ and $gdp_j$ and the trade volume between them ($trade_{ij}$). It is reassuring that the ordering of the average height of the coefficients approximately matches with the order of the correlations reported in Table \ref{covariates}. 
	Note however, that the size of the coefficients is to be interpreted with care because of the limited information available on the unknown claims.
	We also provide prediction intervals in Figure \ref{fig:summary2}, based on the share of real values $x_{ij}$ located in the interval $[q_{0.005},q_{0.955}]$. Here, $q_{0.005}$ and $q_{0.955}$ are the $0.005$ and $0.955$-quantiles derived from the bootstrap distribution (bootstrap sample size $B=100$). On the left, we illustrate the real values against the predicted ones together with grey 95\% prediction intervals for the most recent network. Observations that do not fall within the prediction interval are indicated by red circles. Because of the quadratic mean-variance relation of the exponential distribution it is much easier to capture high values within the prediction intervals than low ones. A circumstance that materializes in the fact that exclusively small values are outside the prediction intervals.
	The share of real values within the prediction intervals against time is shown on the right hand side of Figure \ref{fig:summary2}. We cover on average $96\%$ of all true values with our prediction intervals over all time periods and regard the bootstrap approach therefore as satisfying.
	
	\subsection{Comparison to alternative routines}\label{alternative}
	\noindent \textsl{Gravity model:}
	A standard solution to the problem is the  gravity model (e.g. \citealp{sheldon1998}). In essence it represents a multiplicative independence model
	\begin{equation}
	\label{gravity_multi}
	\hat{\mu}_{ij}=\frac{x_{i\bullet}x_{\bullet j}}{x_{\bullet \bullet}}.
	\end{equation}
	The model is simple, easy to implement and very intuitive. In situations where diagonal elements $x_{ii}$ are not restricted to be zero it even coincides with the maximum entropy solution.\\
	\noindent \textsl{Tomogravity model:}	
	An extension of the gravity model  is given by the tomogravity approach by \citet{zhang2003}. The model was initially designed to estimate point-to-point traffic volumes from dyadic loads and builds on minimizing the loss-function
	\begin{equation}
	\label{tomogravity}
	\hat{\bm{\mu}}=\text{arg} \min \limits_{\bm{\mu}}\bigg{\{} (\mathbf{A}\bm{\mu}-\mathbf{y})^T(\mathbf{A}\bm{\mu}-\mathbf{y})+\psi^2\sum_{i \neq j}\frac{\mu_{ij}}{N}\log\bigg{(}\frac{\mu_{ij}}{x_{i\bullet}x_{\bullet j}}\bigg{)} \bigg{\}}
	\end{equation}
	subject to the non-negativity constraint. Here, the gravity model (\ref{gravity_multi}) serves as a null model in the penalization term and the strength of penalization is given by $\psi$. The approach is implemented in the \texttt{R} package \texttt{tomogravity} (see \citealp{blocker2014}). \citet{zhang2003} show in a simulation study, that $\psi=0.01$ is a reasonable choice if no training set is available.	In our competitive comparison we optimize the tuning parameter in order to minimize the overall $L_2$ error with grid search and find $\psi=0.011$ to be an optimal value.
	\\
	\noindent \textsl{Non-negative LASSO:}	The LASSO (\citealp{tibshirani1996}) was already applied to predict flows in bike sharing networks by \citet{chen2017} and works best with sparse networks. Using this approach, we minimize the loss function
	\begin{equation}
	\label{lasso}
	\hat{\bm{\mu}}=\text{arg} \min \limits_{\bm{\mu}}\bigg{\{}(\mathbf{A}\bm{\mu}-\mathbf{y})^T(\mathbf{A}\bm{\mu}-\mathbf{y})+ \tau \sum_{i \neq j}| \mu_{ij}|\bigg{\}}
	\end{equation}
	where we can drop the absolute value in the penalization term because of the non-negativity constraint (see \citealp{wu2014} for the non-negative LASSO). In order to use the approach in the competitive comparison, we optimize the penalty parameter $\tau$ on a grid for the minimum $L_2$ error and use $\tau=45,483.6$. The models are estimated using the \texttt{R} package \texttt{glmnet} by \citet{friedman2009}.
	\\
	\noindent \textsl{Ecological regression:}	
	In Ecological Inference (see e.g.\ \citealp{klima2016}, \citealp{king2013}), it is often  assumed that the observations at hand are independent realizations of a linear model, parametrized by time-constant transition-shares $\beta_{ij}$. Define the stacked column sums in $t$ by $\mathbf{y}^t_c$ and the stacked row sums in $t$ by $\mathbf{y}^t_r$. Then, the model can be represented as
	\begin{equation}
	\label{ecological}
	\mathbb{E}[\mathbf{Y}^t_r|\mathbf{Y}_c^t=\mathbf{y}_c^t]=\mathbf{B}_c \mathbf{y}_c^t\text{, for }t=1,...,T
	\end{equation}
	where the $(n \times N)$ matrix $\mathbf{B}_c$ contains the  parameters $\beta_{ij}$. In the give form, the problem is not identified for one time period $t$ and it must be assumed that multiple time-points can be interpreted as independent realizations. Additionally, the model is not symmetric, implying that the solution to equation (\ref{ecological}) does not coincide with the solution to
	\begin{equation}
	\label{ecological2}
	\mathbb{E}[\mathbf{Y}^t_c|\mathbf{Y}_r^t=\mathbf{y}_r^t]=\mathbf{B}_r \mathbf{y}_r^t\text{, for }t=1,...,T.
	\end{equation}
	Since, the estimated transition shares are not guaranteed to be non-negative and sum up to one they must be post-processed to fulfil this conditions.  Both models are fitted  via least-squares in \texttt{R}.
	\\
	\noindent \textsl{Hierarchical Bayesian models:}	
	\citet{gandy2017} propose to use simulation-based methods. In their hierarchical models the first step consists of estimating the link probabilities and given that there is a link, the weight is sampled from an exponential distribution: 
	\begin{equation}
	\label{erdos}
	\begin{split}
	P(X_{ij}>0)&=p\\
	X_{ij}|X_{ij}>0&\sim Exp(\mu).
	\end{split}
	\end{equation}
	In order to estimate the link probabilities $p$, knowledge of the density or a desired  target density is needed. In their basic model it is proposed to use an Erd\"os-R\'enyi model with $p$  consistent with the target density. In an extension of the model, inspired by Graphon models, \citet{gandy2018} propose a so called empirical fitness model. Here the link probability is determined by the logistic function
	\begin{equation}
	\label{fitness}
	P(X_{ij}>0)=p_{ij}=\frac{1}{1+\exp(-\alpha-z_i-z_j)},
	\end{equation}
	with $\alpha$ being some constant that is estimated for consistency with the target density. For the fitness variables $z_i$, the authors propose to use an empirical Bayes approach, incorporating the information of the row and column sums as $z_i=\log(x_{\bullet i}+x_{i \bullet})$.  An implementation of both models is given by the \texttt{R} package \texttt{systemicrisk}. In order to make the approach as competitive as possible we use for each quarter the real (but in principle unknown) density of the networks. Because the results of the method differ between each individual estimate, we average the estimates and evaluate the combined dataset. 
	\subsection{Competitive comparison}
	\begin{table}[t!]
		\resizebox{\textwidth}{!}{
			\begin{tabular}{clccccccc} 
				\hline 
				&	Method& Model & overall $L_1$ & overall $L_2$ & average $L_1$ & SE & average $L_2$ & SE\\
				\hline \hline  
				1&	Regression ($gdp_i$, $gdp_j$, $trade_{ij}$) & (\ref{log_lin_mean2})&{\bf3\,300.242}&   {\bf56.794}&   {\bf  63.466}&    {\bf 9.246}&    {\bf 7.802}&    {\bf 1.085} \\ 
				
				2&	Gravity model&(\ref{gravity_multi})&$4\,300.927$ & $79.342$ & $82.710$ & $13.003$ & $10.935$ & $1.232$ \\ 
				3&	Tomogravity Model & (\ref{tomogravity})&$4\,241.299$ & $75.760$ & $81.563$ & $12.774$ & $10.442$ & $1.168$ \\ 			4&		Non-negative LASSO & (\ref{lasso})&$7\,233.821$ & $127.638$ & $139.112$ & $20.399$ & $17.572$ & $2.150$ \\
				5&Ecological Regression, columns & (\ref{ecological})&	$9\,785.014$ & $163.422$ & $188.173$ & $24.575$ & $22.573$ & $2.032$ \\ 
				6&Ecological Regression, rows & (\ref{ecological2})	&$10\,776.570$ & $184.636$ & $207.242$ & $27.662$ & $25.438$ & $2.946$\\ 
				7&Hierarchical,  Erd\"os-R\'enyi& (\ref{erdos})&$5\,328.834$&  $101.639$ & $102.478$&   $17.439$&   $14.004$&    $1.610$\\ 
				8&Hierarchical, Fitness& (\ref{fitness})&$5\,316.036$&  $102.072$&  $102.231$&   $17.912$&   $14.039$&    $1.827$\\ 
				\hline \\[-1.8ex] 
			\end{tabular} 
		}
		\caption{Comparison of different methods with the BIS Dataset. All values scaled by $100\,000$  and lowest values in bold.} 
		\label{comparison2} 
	\end{table} 
	Again we compare the different algorithms in terms of their $L_1$ and $L_2$ errors in Table \ref{comparison2}. In the first row of Table \ref{comparison2} we show the restricted maximum likelihood model with the best predictions from Table \ref{comparison} and in the following rows, the models introduced in Section \ref{alternative} above are shown. The results  can be separated roughly into three blocks. The models that fundamentally build on some kind of Least Squares criterion without referring in some way to the gravity model or the maximum entropy solution (ecological regression, and non-negative LASSO in rows four, five and six) have the highest values in terms of their $L_1$ and $L_2$ errors.  
	Somewhat better are the Hierarchical Bayesian Models (rows seven and eight) that can be considered as the second block. However, although they provide better predictions than the models in the first block, we used the real density of the network in order to calibrate them which gives them an unrealistic advantage. The third group is given by the gravity and tomogravity model (rows two and three). Those are statistically indistinguishable and provide considerably better results than the models from the former blocks. Nevertheless, the regression model that uses exogenous information on $trade_{ij}$ (first row) yields the best predictions in this comparison.
	
	\subsection{Performance of the estimator} \label{perform}
	We hope to see improvements in the predictions if we include informative exogenous variables in the model. Informative means in this context, that variation in $\tilde{\mathbf{z}}_{ij}$ is able to explain variation in the unknown $X_{ij}$. Apparently, including information with a low association to $X_{ij}$ simply adds noise into the estimation procedure. In this case we expect inferior predictions as compared to the IPFP solution. We illustrate the properties of the estimator using a simulation study with the following data generating process 
	\begin{equation}
	\label{DGP1}
	\begin{split}
	\delta_i&\sim N(0,1)\text{, }\gamma_j\sim N(0,1)\text{, }\tilde{\mathbf{z}}_{ij}\sim N(0,1)\text{, for }i, j=1,...,10 \text{ and }i\neq j \\
	\mu_{ij}(\beta)&=\exp(\delta_i + \gamma_j + \tilde{\mathbf{z}}_{ij}\beta)\\
	X_{ij}&\sim Exp(\mu_{ij}(\beta)).
	\end{split}
	\end{equation} 
	Since the association between $\tilde{\mathbf{z}}_{ij}$ and the unknown $X_{ij}$ is crucial, we vary the parameter $\beta$ from $-4$ to $4$ and denote with $\mu_{ij}(\beta)$ the mean based on $\beta$. For each parameter $\beta$ we re-run the data generating process (\ref{DGP1}) $S=1\,000$ times and calculate for the $s$-th simulation the IPFP solution $\check{\mu}_{s,ij}(\beta)$ and the restricted maximum likelihood solution $\hat{\mu}_{s,ij}(\beta)$. Based on that, we calculate in each simulation the ratio of the  squared errors 
	\begin{equation*}
	RSS_{s}(\beta)=\frac{\sum_{i\neq j}(X_{s,ij}-\check{\mu}_{s,ij}(\beta))^2}{\sum_{i\neq j}(X_{s,ij}-\hat{\mu}_{s,ij}(\beta))^2}\text{, for }s=1,...,1000 \\.
	\end{equation*}
	This ratio is smaller than one if the IPFP estimates yield a lower mean squared error than the restricted maximum likelihood estimates and higher than one if the exogenous information improves the predictive quality in the terms of the mean squared error.
	\begin{figure}[t!]
		\includegraphics[trim={0cm 0cm 0cm 0cm},clip,width=1\textwidth]{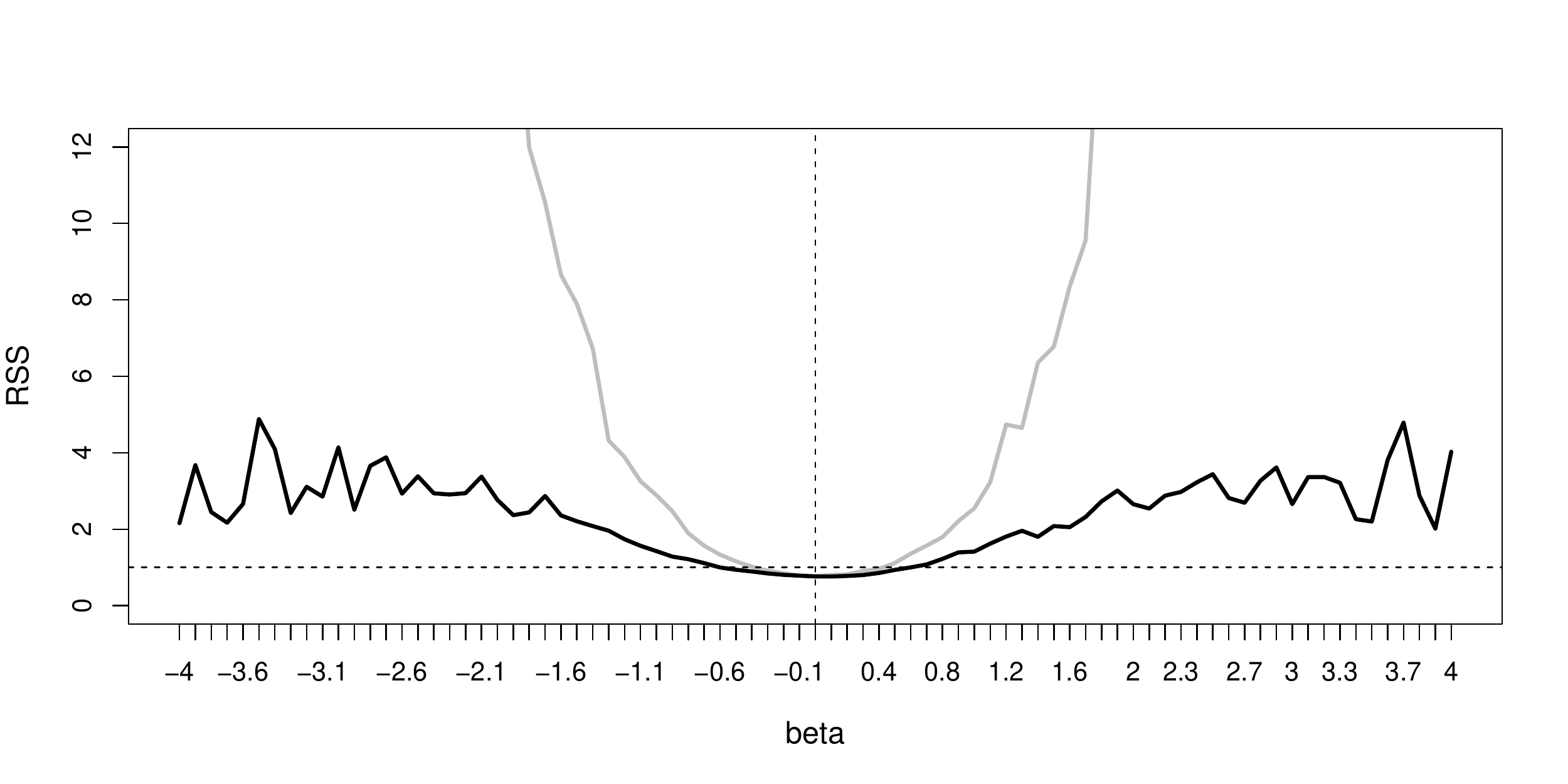}
		
		\caption{Median (solid black) and mean (solid grey) of the relative squared error $RSS_{s}(\beta)$  for different levels of $\beta$.}
		\label{fig:compare}
	\end{figure}
	
	In Figure \ref{fig:compare}, we show the median (solid black) and the mean (solid grey) of $RSS_{s}(\beta)$ for different values of $\beta$ as well as a horizontal line indicating the value one (dashed black) and a vertical line for $\beta=0$ (dashed black). It can be seen, that the mean and the median of $RSS_{s}(\beta)$ are below one for values of $\beta$ that are roughly between $-0.5$ and $0.5$ but increase strongly with higher absolute values of $\beta$. Apparently, the distribution of $RSS_{s}(\beta)$ is skewed with a long tail since the mean is much higher than the median. With very high or low values of $\beta$, the median of the relative mean squared error becomes more volatile and partly decreases. 
	\begin{figure}[htpb]
		\centering
		\begin{subfigure}{\textwidth}
			\centering	\includegraphics[trim={0cm 1.5cm 0cm 0cm},clip,width=0.9\textwidth]{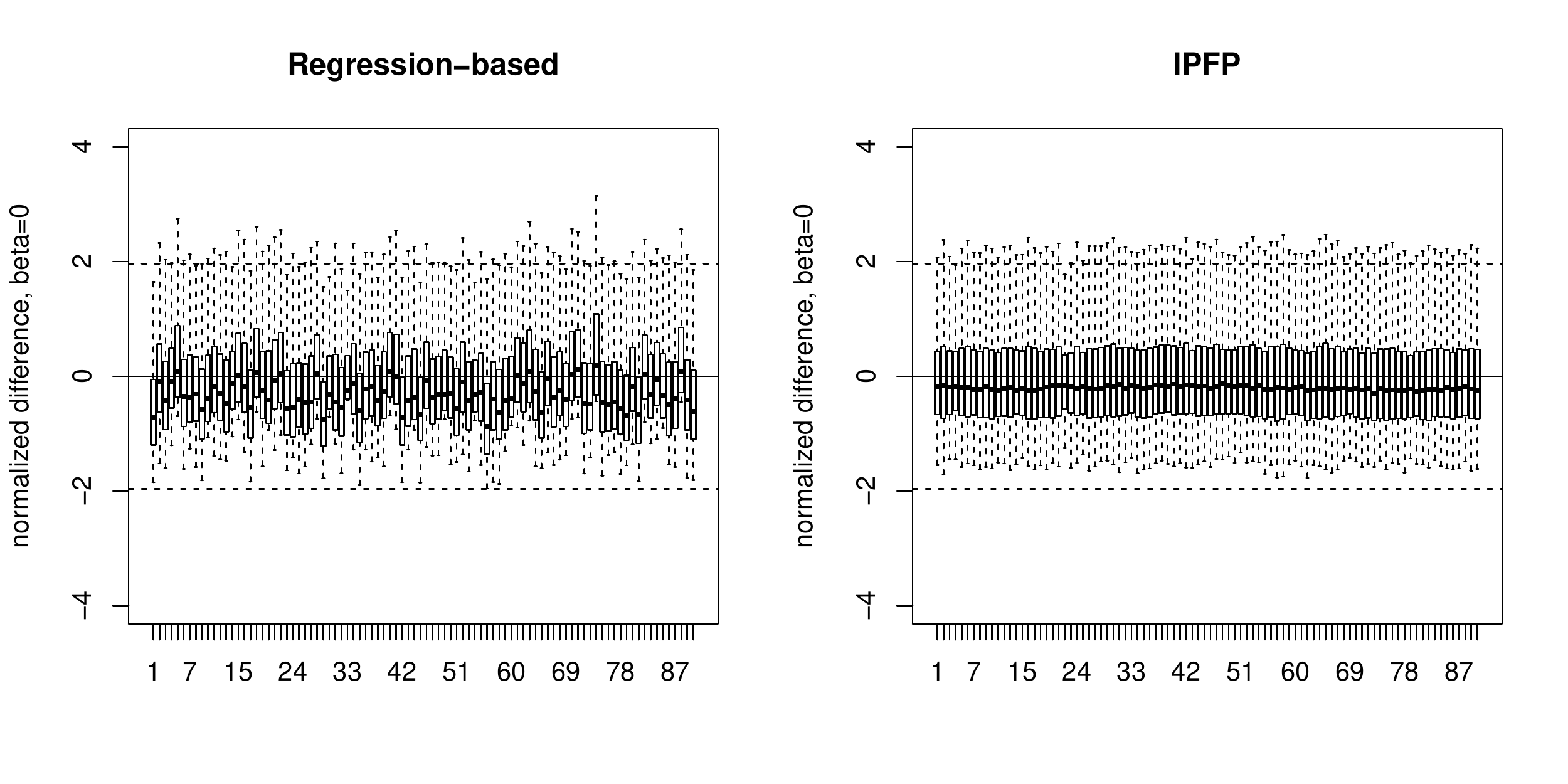}
		\end{subfigure}
		\begin{subfigure}{\textwidth}
			\centering	\includegraphics[trim={0cm 1.5cm 0cm 1.8cm},clip,width=0.9\textwidth]{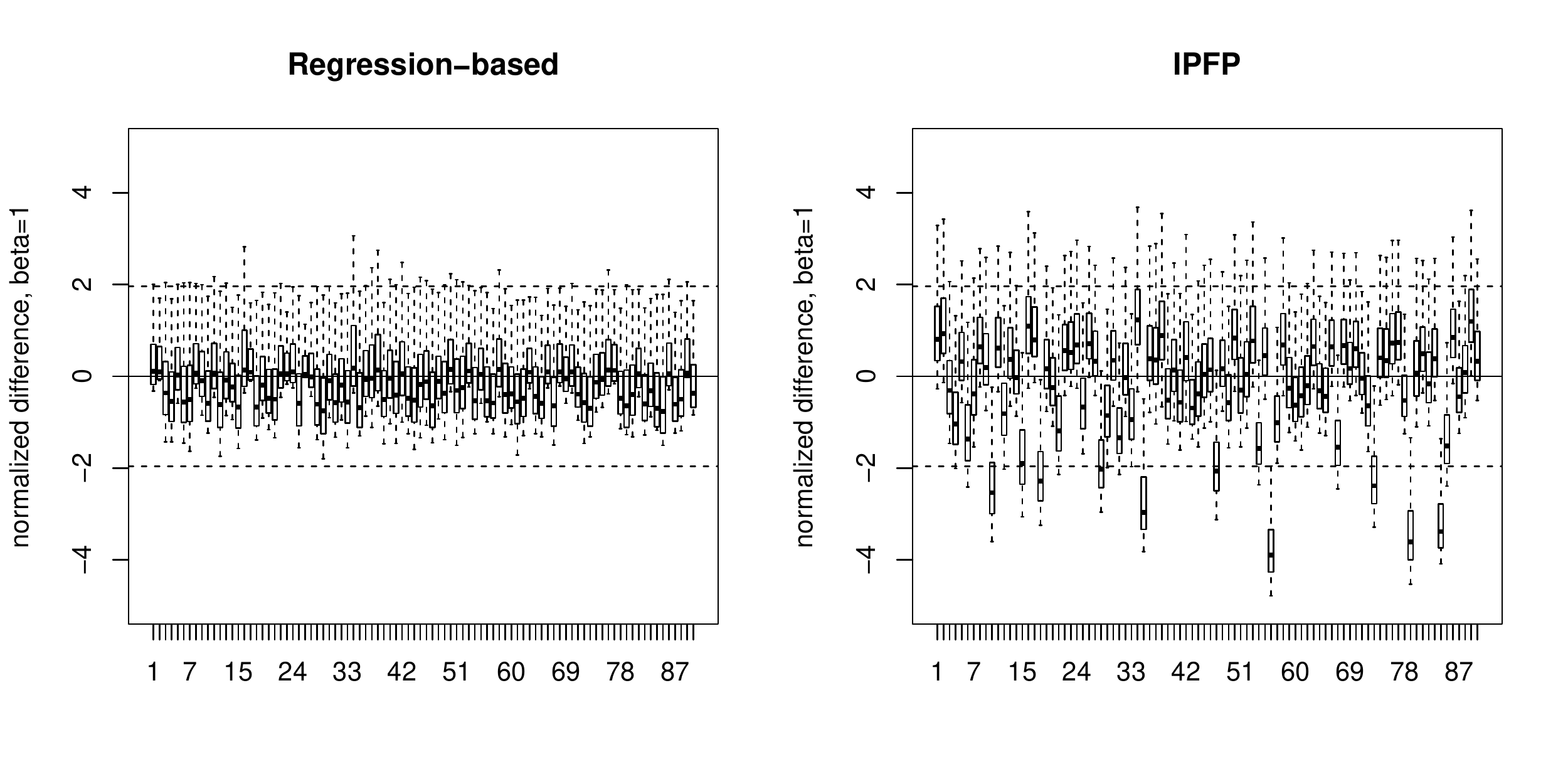}
		\end{subfigure}
		\begin{subfigure}{\textwidth}
			\centering	\includegraphics[trim={0cm 0.5cm 0cm 1.8cm},clip,width=0.9\textwidth]{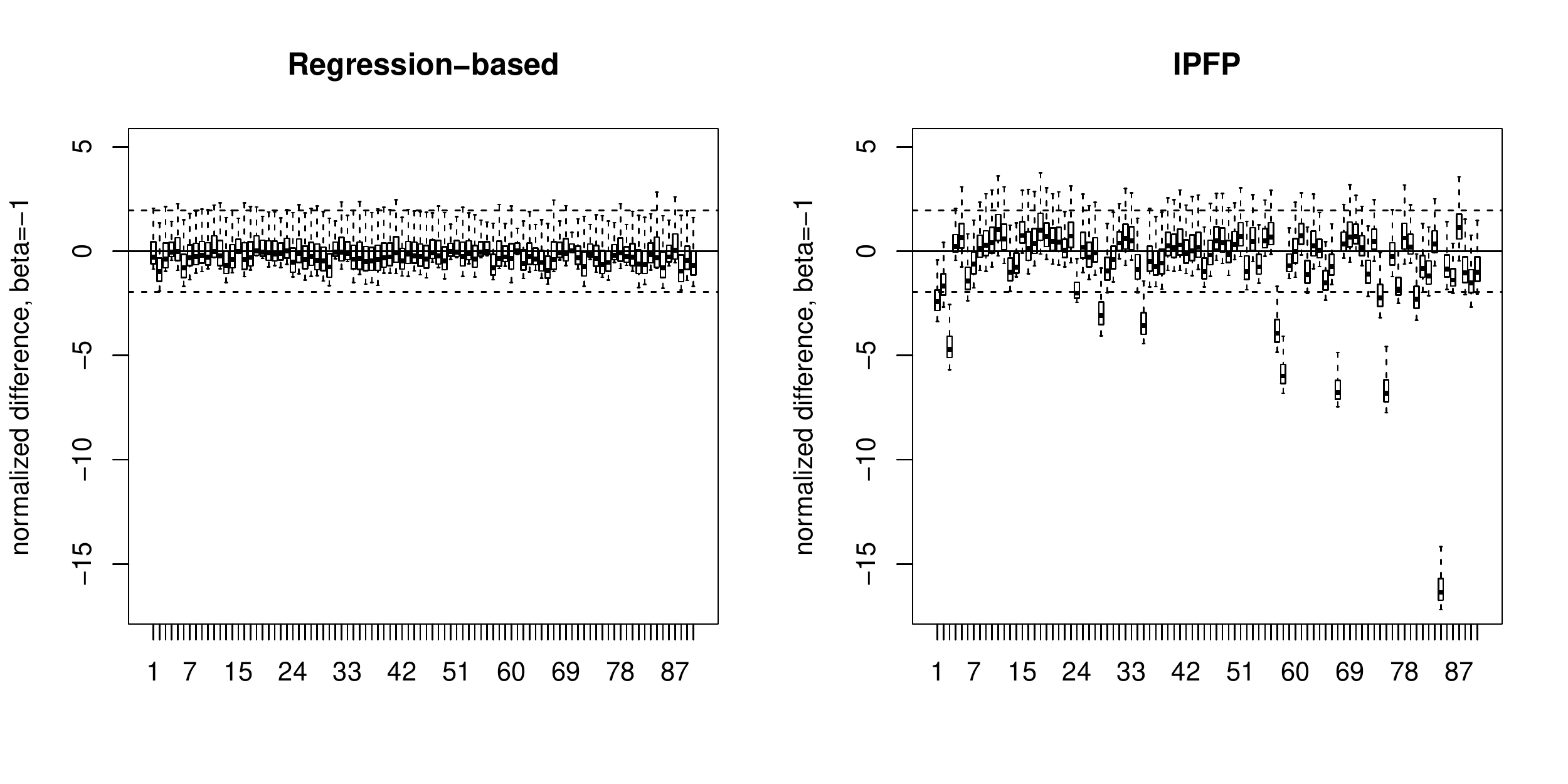}
		\end{subfigure}
		\caption{Boxplots for the normalized differences between the true expectation and the mean of the estimated ones $\Delta_{s,ij}(\beta)$ for $\beta=0$ (top), $\beta=1$ (middle) and $\beta=-1$ (bottom). Regression-based model on the left, IPFP on the right. Outliers are excluded from the boxplot representation for better clarity.}
		\label{fig:compare2}
	\end{figure}
	
	Furthermore, we investigate how well the estimated expectations match the true ones. By construction, it holds that $\mathbf{A}\bm{\mu}=\mathbb{E}[\mathbf{Y}]=\mathbf{A}\mathbb{E}[\hat{\bm{\mu}}]$ and consequently, the regression-based approach as well as IPFP assume that the sum of realized values equals the sum of the true expectations. Nevertheless, the moment condition does not  imply that $\mathbb{E}[\hat{\mu}_{s,ij}]=\mu_{ij}$. In order to investigate potential bias, we draw again from 
	\begin{equation*}
	\label{DGP2}
	\begin{split}
	\delta_i&\sim N(0,1)\text{, }\gamma_j\sim N(0,1)\text{, }\tilde{\mathbf{z}}_{ij}\sim N(0,1)\text{, for }i, j=1,...,10 \text{ and }i\neq j,
	\end{split}
	\end{equation*} 
	and fix $\mu_{ij}(\beta)=\exp(\delta_i + \gamma_j + \tilde{\mathbf{z}}_{ij}\beta)$ to be the true expectation and draw and re-estimate again $S=1\,000$ times from $X_{ij}\sim Exp(\mu_{ij}(\beta))$. Apparently, estimating the true expectations is a hard task as only sums of random variables with different expectations are available. Consequently, the variation of the mean estimates is rather high and we report boxplots of the normalized difference between the true value and mean estimate
	\begin{equation*}
	\Delta_{s,ij}(\beta)=\frac{\hat{\mu}(\beta)_{s,ij}-\mu_{ij}(\beta)}{S^{-1}\sum_{s=1}^{S}(\hat{\mu}(\beta)_{s,ij}-S^{-1}\sum_{s=1}^{S}\hat{\mu}(\beta)_{s,ij})^2},
	\end{equation*}
	and accordingly for $\check{\mu}(\beta)_{s,ij}$.
	In Figure \ref{fig:compare2} we illustrate three different cases with $\beta=0$ (top), $\beta=1$ (middle) and $\beta=-1$ (bottom). On the left hand side, boxplots for $\Delta(\beta)_{s,ij}$ are shown for the regression-based model and on the right hand side for IPFP. The solid black line represents zero and the dashed black lines give $\pm$ 1.96. The results for the case $\beta=0$ on the top, match with the previous analysis illustrated in Figure \ref{fig:compare} and show that IPFP identifies the true expectations somewhat better than the regression-based approach when the exogenous information is non-informative. In such a case, including $\tilde{\mathbf{z}}_{ij}$ adds noise in the estimation procedure, resulting in a greater variance around the true expectations. However, this changes strongly if $\tilde{\mathbf{z}}_{ij}$ is informative. Especially for $\beta=-1$ on the bottom of Figure \ref{fig:compare2}, some estimates obtained from IPFP are seriously biased because this procedure does not have the ability to account for the dyad-specific heterogeneity. The regression-based method, however does a reasonable job in recovering the unknown true expectations.
	\FloatBarrier
	\section{Discussion}\label{conc}
	In this paper we propose a method that allows for network reconstruction within a regression framework. This approach makes it easy to add and interpret exogenous information. It also allows to construct bootstrap prediction intervals to quantify the uncertainty of the estimates.
	Furthermore, the framework is flexible enough to deal with problems that involve partial information. For example if some elements of the network are known or if we have information on the binary network, then we can model the expected values of the matrix entries conditional this information, simply by changing the routing matrix and the E-Step.
	
	However, we also want to list some shortfalls of the method. An obvious drawback of the method is its derivation from the maximum entropy principle that tries to allocate the matrix entries as even as possible and is therefore not very suitable for sparse networks as long as the sparseness cannot be inferred from the marginals.  Furthermore, the estimated coefficients must be interpreted with care, as they are estimated based on a data situation with much less information than in usual regression settings.
	As a last but most important point, if the association between the exogenous explanatory variable(s) and the unknown matrix entries is low, the method is likely to deliver predictions that are \textsl{worse} than simple IPFP. It is therefore highly recommendable to use expert knowledge when selecting the exogenous dyadic covariates for regression-based network reconstruction.
	\FloatBarrier
\newpage
%
%
%

	\bibliographystyle{Chicago}
	
	\bibliography{literature}
	\newpage
	\pagenumbering{Roman}
\appendix
\section{Estimation with random effects}\label{raneff}
In order to fit a model of the form
\begin{equation*}
\begin{split}
\mu_{ij}(\bm{\theta})&=\exp(\delta_i + \gamma_j + \tilde{\mathbf{z}}_{ij}^T\bm{\beta})=\exp(\mathbf{z}_{ij}^T\bm{\theta}), \\ 
\begin{pmatrix}
\delta_i\\
\gamma_j
\end{pmatrix}&\sim \mathcal{N}_2\left(\mathbf{0},\begin{pmatrix}
\sigma^2_\delta & \sigma^2_{\delta,\gamma}   \\
\sigma^2_{\delta,\gamma} & \sigma^2_{\gamma}  
\end{pmatrix}\right)\text{, for }i,j=1,...,n\text{ and }i\neq j,
\end{split}
\end{equation*}
we follow a Laplace approximation estimation strategy similar to \citet{breslow1993} and fix  $\bm{\vartheta}$ to some value $\bm{\vartheta}_0$, see also \citet{tutz2010}.
The moment condition implies a restriction for $\bm{\theta}$ but not for $\bm{\vartheta}$ and given some starting value $\bm{\theta}_0$, we can maximize the \textsl{penalized log-likelihood} (constant terms omitted) 
\begin{equation*}
\ell_{pen}(\bm{\theta};\bm{\vartheta}_0,\bm{\theta}_0)=	\sum_{q\in \mathcal{I}}\bigg{(} -\mathbf{z}_q^T\bm{\theta} -\exp\{\mathbf{z}_q^T(\bm{\theta}_0-\bm{\theta}) \}  \bigg{)}-\frac{1}{2}(\bm{\delta}^T,\bm{\gamma}^T)\bm{\Sigma}^{-1}(\bm{\vartheta}_0)(\bm{\delta}^T,\bm{\gamma}^T)^T
\end{equation*}
subject to the moment condition $\mathbf{A}\bm{\mu(\bm{\theta})}=\mathbf{y}$. Therefore, the new optimization problem is given by
\begin{equation}
\label{auglag2}
\mathcal{L}(\bm{\theta};\bm{\xi},\zeta,\bm{\theta}_0,\bm{\vartheta}_0) = -\ell_{pen}(\bm{\theta};\bm{\vartheta}_0,\bm{\theta}_0)-\bm{\xi}^T(\mathbf{A}\bm{\mu}(\bm{\theta}) -\mathbf{y}  ) + \frac{\zeta}{2} ||\mathbf{A}\bm{\mu}(\bm{\theta}) -\mathbf{y}    ||^2_2.
\end{equation}
Define $\tilde{\mathbf{Z}}$ as the $(N \times l)$ design matrix for the fixed effects and $\mathbf{U}$ as the ($N \times 2n$) random effects design matrix, this allows to write the generic mean as $\log(\bm{\mu})=\mathbf{Z}\bm{\theta}=\tilde{\mathbf{Z}}\bm{\beta} + \mathbf{U}(\bm{\delta}^T,\bm{\gamma}^T)^T$. Given that we have some estimate of $\bm{\theta}$, call it $\bm{\theta}_1$ we can estimate the variance parameters $\bm{\vartheta}$ with an approximation of the \textsl{marginal restricted log-likelihood}:
\begin{equation}
\label{pen1}
\ell_R(\bm{\vartheta};\bm{\theta}_1,\bm{\theta}_0)=-\frac{1}{2}\log(|\mathbf{V}(\bm{\vartheta})|)-\frac{1}{2}\log(|\tilde{\mathbf{Z}}^T\mathbf{V}(\bm{\vartheta})^{-1}\tilde{\mathbf{Z}}|)-\frac{1}{2}(\tilde{\mathbf{y}}-\tilde{\mathbf{Z}}\bm{\beta)}^T\mathbf{V}(\bm{\vartheta})^{-1}(\tilde{\mathbf{y}}-\tilde{\mathbf{Z}}\bm{\beta})
\end{equation}
where $\mathbf{V}(\bm{\vartheta})=(\mathbf{D} \text{diag}\{\mathbf{V}(\mathbf{X}) \}^{-1}\mathbf{D})^{-1}+\mathbf{U}\bm{\Sigma}(\bm{\vartheta})\mathbf{U}^T$, with $\mathbf{D}=\text{diag}\{\bm{\mu}(\bm{\theta}_1)\}$ and $\text{diag}\{\mathbf{V}(\mathbf{X}) \}^{-1}=\text{diag}\{\bm{\mu}(\bm{\theta}_1)^{-2}\}$ and consequently $\mathbf{V}(\bm{\vartheta})=\mathbf{I}_{N}+\mathbf{U}\bm{\Sigma}(\bm{\vartheta})\mathbf{U}^T$. The pseudo-observations $\tilde{\mathbf{y}}$ are given by  $\log(\bm{\mu}(\bm{\theta}_1))+\mathbf{D}^{-1}(\bm{\mu}(\bm{\theta}_0)-\bm{\mu}(\bm{\theta}_1))$. Estimators can be obtained by iteratively optimizing firstly  (\ref{auglag2}) and secondly  (\ref{pen1}) in each iteration until convergence.

\end{document}